\documentclass{nature}
\usepackage{graphicx}
\makeatletter
\let\saved@includegraphics\includegraphics
\AtBeginDocument{\let\includegraphics\saved@includegraphics}
\renewenvironment*{figure}{\@float{figure}}{\end@float}
\makeatother
\usepackage[square,sort,comma,numbers]{natbib}
\usepackage{amsmath, multirow,amssymb}
\usepackage{hyperref,breakurl}
\newcommand\ion[2]{#1$\;${%
\ifx\@currsize\normalsize\small \else
\ifx\@currsize\small\footnotesize \else
\ifx\@currsize\footnotesize\scriptsize \else
\ifx\@currsize\scriptsize\tiny \else
\ifx\@currsize\large\normalsize \else
\ifx\@currsize\Large\large
\fi\fi\fi\fi\fi\fi
\rmfamily\@Roman{#2}}\relax}%

\def\HST{{\it HST}}
\def\JWST{{\it JWST}}
\def\Spitzer{\it Spitzer}

\def\chone{3.6\:\mu\rm{m}}
\def\chtwo{4.5\:\mu\rm{m}}

\def\macsb{MACS~J1149.5$+$2223}
\newcommand{\surfsup}{SURFS UP}

\newcommand{\hb}{\mbox{H$\beta$}}

\newcommand{\nii}{N~{\sc ii}}
\newcommand{\oiii}{[O~{\sc iii}]}
\newcommand{\oiiif}{[O~{\sc iii}]4959, 5007$\mbox{\AA}$}

\def\otf{[\ion{O}{3}] 88$~\mu\mbox{m}$}

\def\lya{Lyman-$\alpha$}
\def\ha{H$\alpha$}
\newcommand{\hst}{\it HST}

\title{The High-redshift Universe with Spitzer}

\author{Maru\v{s}a Brada\v{c}$^{1,2}$}

\begin{document}
\maketitle
\begin{affiliations}
 \item  Department of Physics, University of California, Davis, CA 95616, USA
\end{affiliations}

\begin{abstract}  
  When did galaxies start forming stars? What is the role of distant galaxies in galaxy formation models and the epoch of reionization?  What are the conditions in typical star-forming galaxies at $z \gtrsim 4$? Why is galaxy evolution dependent on environment?  The {\Spitzer} Space Telescope has been a crucial tool for addressing these questions. Accurate knowledge of stellar masses, ages, and star formation rates (SFRs) requires measuring rest-frame optical (and UV) light, which only {\Spitzer} can probe at high-redshift for a sufficiently large sample of typical galaxies. Many of these science goals are the main science drivers for James Webb Space Telescope, and {\Spitzer} afforded us their first exploration.
\end{abstract}

Very little was known about the high-redshift universe when the
{\it Spitzer Space Telescope} \citep{werner04}
was being planned in the 90's. The then optimistic science cases that were put forward
were predicting Spitzer could observe ordinary galaxies at $z\sim 5$
[\citealp{sbook}, Chapter 15]. Given that the highest-redshift galaxy discovered
at that time was $z_{\rm spec}=4.7$ [\citealp{schneider89}], this was quite an extraordinary
claim. It was not until 1998 that the redshift five barrier was
broken \citep{dey98}. While spectroscopic searches with {\lya} were
proposed in the 1960's, and radio galaxies were discovered beyond $z>3$
in the late 1980's \citep{lilly88,chambers90},  it was the  powerful new
photometric  method \citep{steidel96}  to search
for ``Lyman Break Galaxies'' (LBGs) that opened a
new window for efficient discoveries of high-redshift galaxies. This
 meant that high redshift galaxies were being
discovered in large numbers, in particular with \emph{Hubble} Space
Telescope ({\hst}). 

\begin{figure}
\includegraphics[width=\textwidth]{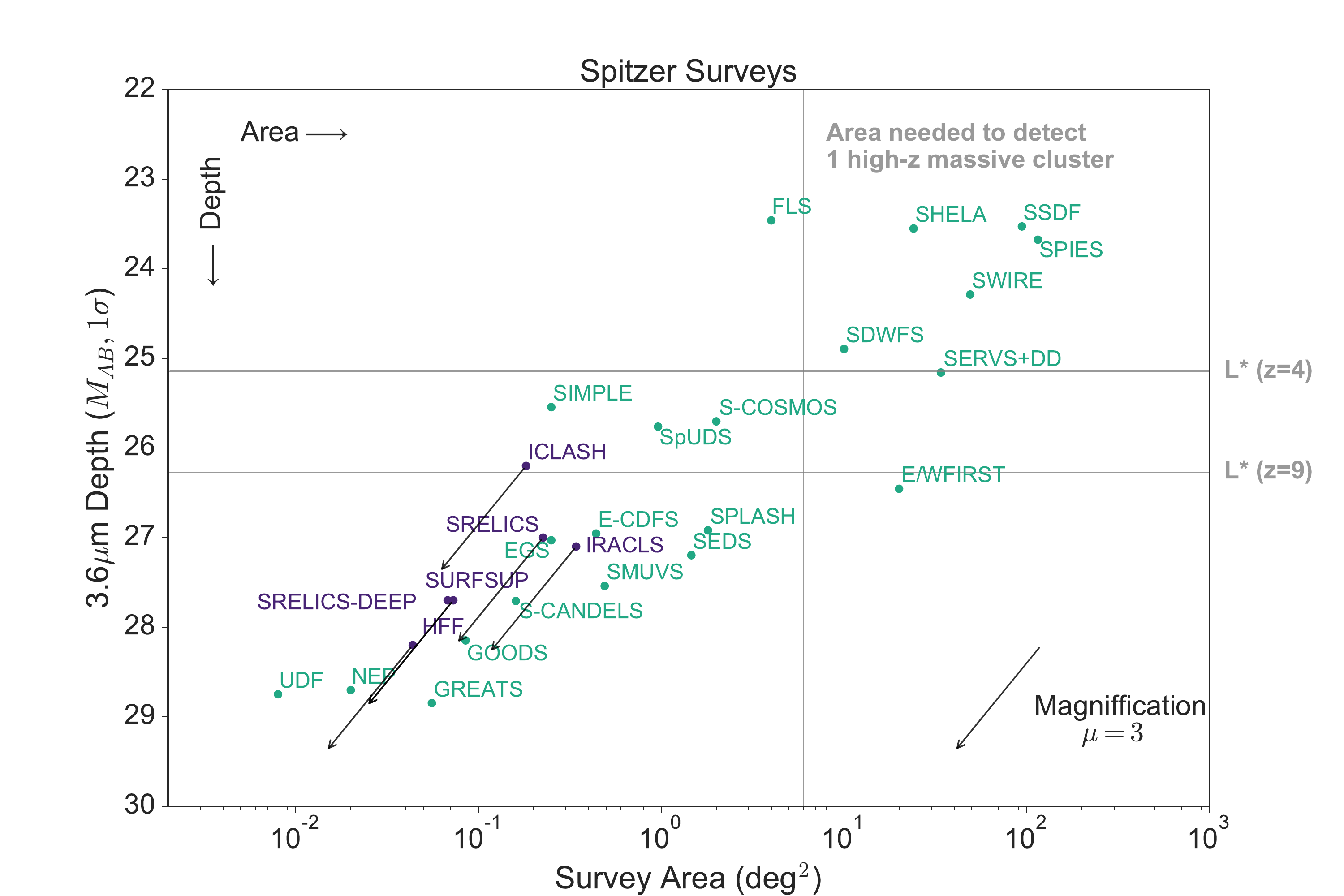}
  \caption{{\bf Sensitivities and area of Spitzer Surveys.} The ${\chone}$ 1-$\sigma$ point-source sensitivities vs area
    surveyed for
    surveys executed during the {\Spitzer} mission. Recent
    {\Spitzer} Surveys have reached the depths that allow us to push
    observations to observe typical ($L^{\ast}$) galaxies at e.g., $z>4$ and
    $z>9$
    (horizontal grey lines). Plotted are depths and areas for both field (green) and lensing (purple)
    surveys as symbols. The lensing surveys have a typical magnifications of
    $\mu \sim 3$, the effect of such a magnification (increased depth and
    reduced area) are indicated with arrows. The typical area needed to
    detect $\sim 1$ high-redshift ($z\gtrsim 1$) massive cluster
    ($M>10^{14}M_{\odot}$) is indicated by the vertical grey line.
Adapted from \citep{ashby18}. \label{fig:surveys}}

\end{figure}

Spitzer, after its launch, soon followed these discoveries. While
ordinary galaxies at $z\gtrsim 5$ (i.e., those with characteristic
luminosities $L^*$) needed large time investment before they could be
detected, there are now many surveys that are deep enough to observe
those galaxies and study their properties
(Fig.~\ref{fig:surveys}). Among the first studies of $z> 5$ galaxies
with Spitzer were observations with Spitzer's Infrared Array
Camera (IRAC, \citealp{fazio04}) of objects gravitationally lensed by
foreground clusters. This allowed early detection of galaxies at
$z=6.7$ [\citealp{egami05}] and $z=6.56$ [\citealp{chary05}].  This
was only possible through the considerable magnification of these
galaxies afforded by the presence of the massive cluster.
Gravitational lensing  resulted in a significant increase
in depth and resolution, a feature that {\Spitzer} has continued to benefit from
ever since.  Larger samples at $z\gtrsim 5$ were discovered
concurrently by virtue of  deep observations of the HUDF \citep{yan05, eyles05}. The
exposure time needed to reach the required depth was many hours and for
the first time this was achieved with {\Spitzer} Great Observatories
Origins Deep Survey (GOODS, PI Dickinson) program
for a large enough area.

The present landscape has
changed considerably since early {\Spitzer} observations. While {\hst} is still the prime telescope to
search for and identify the highest redshift galaxies,  {\Spitzer}  has allowed us
to study these galaxies in detail. For cosmic dawn sources (i.e., galaxies
at $z>6$) {\Spitzer} became a key telescope not
only to help establish redshifts, but also in the
study of stellar masses, star formation rates, optical emission line
strengths, and the identification of old stellar populations.  

For intermediate redshift sources ($4<z<6$ in this review), {\Spitzer}
has also been crucial in robustly determining their stellar masses. In
addition, although not discussed in detail here (but see Comment to
Nature Astronomy by Daniela Calzetti), at $z < 2$ observations with
the Multiband
  Imaging Photometer for {\Spitzer} (MIPS, \citealp{rieke04}) have been central to measurements of cosmic star formation history.  Finally, {\Spitzer} was not
  only succesful in finding high redshift galaxy clusters ($z>1$),
  but also in studying the stellar properties of their cluster members. 
In this review, we give a brief overview of
how {\Spitzer} has provided measurements confirming our expectations, as
well as of new puzzles that are changing the paradigm of galaxy formation at
high redshifts.

\section{Cosmic Dawn}
\label{sec:cosmicdawn}

Tracing star formation to the earliest times has been a long-standing
goal of extragalactic astronomy. In particular, studying the onset of
star formation is of importance not only for galaxy formation models,
but also for  studies of the early
universe. The cosmic Dark Ages - when the Universe was filled with
neutral hydrogen that was opaque to ultraviolet 
(UV) light - are thought to have ended around 500 million years
after the Big Bang, when early light sources produced enough energetic
photons to ionize the neutral hydrogen \citep{planckXIII}. This phase is referred to as the epoch of reionization, and is also the era of the formation of the first
galaxies. It is now clear that the process was completed by
$z \sim 6$ [\citealp{fan06,robertson15,mason19}]. However, a direct link between
early light sources and reionization requires
a detailed understanding of when and how galaxies first formed and
built up their stellar content.

\begin{figure}
\includegraphics[width=\textwidth]{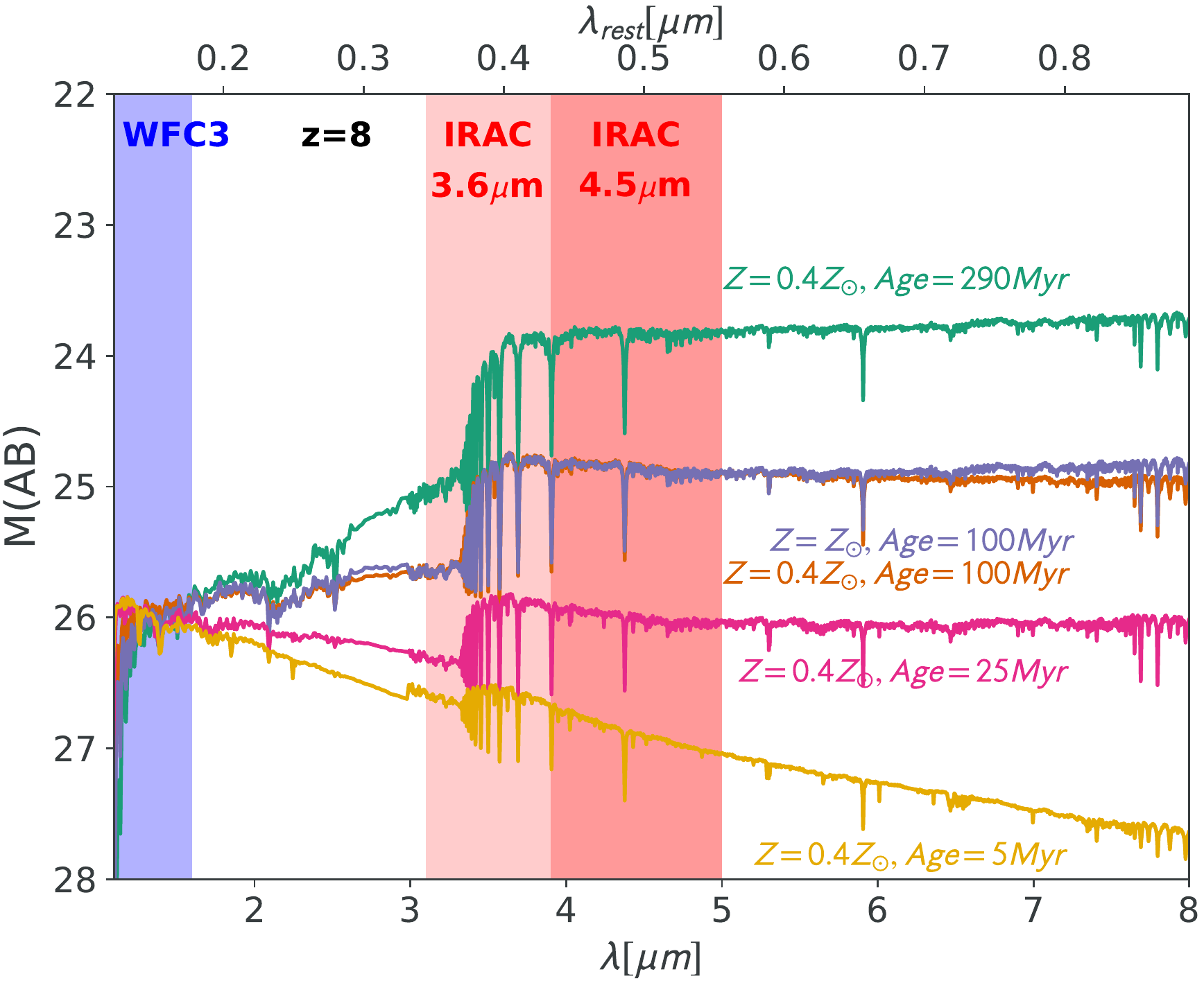} 
  \caption{{\bf Spitzer data is crucial for determining stellar
     ages.} Five different spectral energy distributions (SEDs) for starburst galaxies of
    different ages (from \citealp{bc03}) redshifted to $z=8$.
    Whereas all these galaxies have similar colors in HST/WFC3
    bands (blue shaded region) relative to typical photometric uncertainties, the different ages can be easily
    distinguished once ${\chone}$ and ${\chtwo}$ {\it Spitzer}/IRAC imaging
    is added (the  ${\chone}$ and ${\chtwo}$ IRAC bands are shown with the light and dark red shaded regions), as their
    $m_{H_{\rm{160W}}}-m_{\chone}$ and $m_{H_{\rm{160W}}}-m_{\chtwo}$
    colors are very different. We plot a $Z=0.4Z_{\odot}$ starburst
    galaxy SED with stellar population at $t=290\mbox{ Myr}$ after the
    burst (green), $100\mbox{ Myr}$ (orange), $25\mbox{ Myr}$ (pink) and
    $5\mbox{ Myr}$ (yellow). Also shown is a $t=100\mbox{ Myr}$ SED
    with $Z=Z_{\odot}$ (purple), to show the effect of
    metallicity degeneracy with age, which is small. \label{fig:sed}}
\end{figure}

{\it Spitzer} played (and is continuing to do so with the archival data)
a unique role in advancing our understanding of the formation and
evolution of galaxies at $z\gtrsim 6$. Deep observations at ${\chone}$
and ${\chtwo}$ with IRAC probe rest-frame optical properties of these
galaxies, hence {\Spitzer} data is critical for age (Fig.~\ref{fig:sed})
and stellar mass determination at high-redshift. While observations
with {\HST} measure UV light emitted by young stars, {\it Spitzer}
IRAC measures the rest-frame optical light from long-lived stars in
galaxies.  In addition, light from some of the most prominent rest-frame
optical emission lines (e.g., {\oiiif}+{\hb}) enter Spitzer ${\chone}$
and ${\chtwo}$ bands at $z>6$, allowing us to measure their
contributions.

\begin{figure}
\begin{minipage}{\textwidth}
\includegraphics[width=\textwidth]{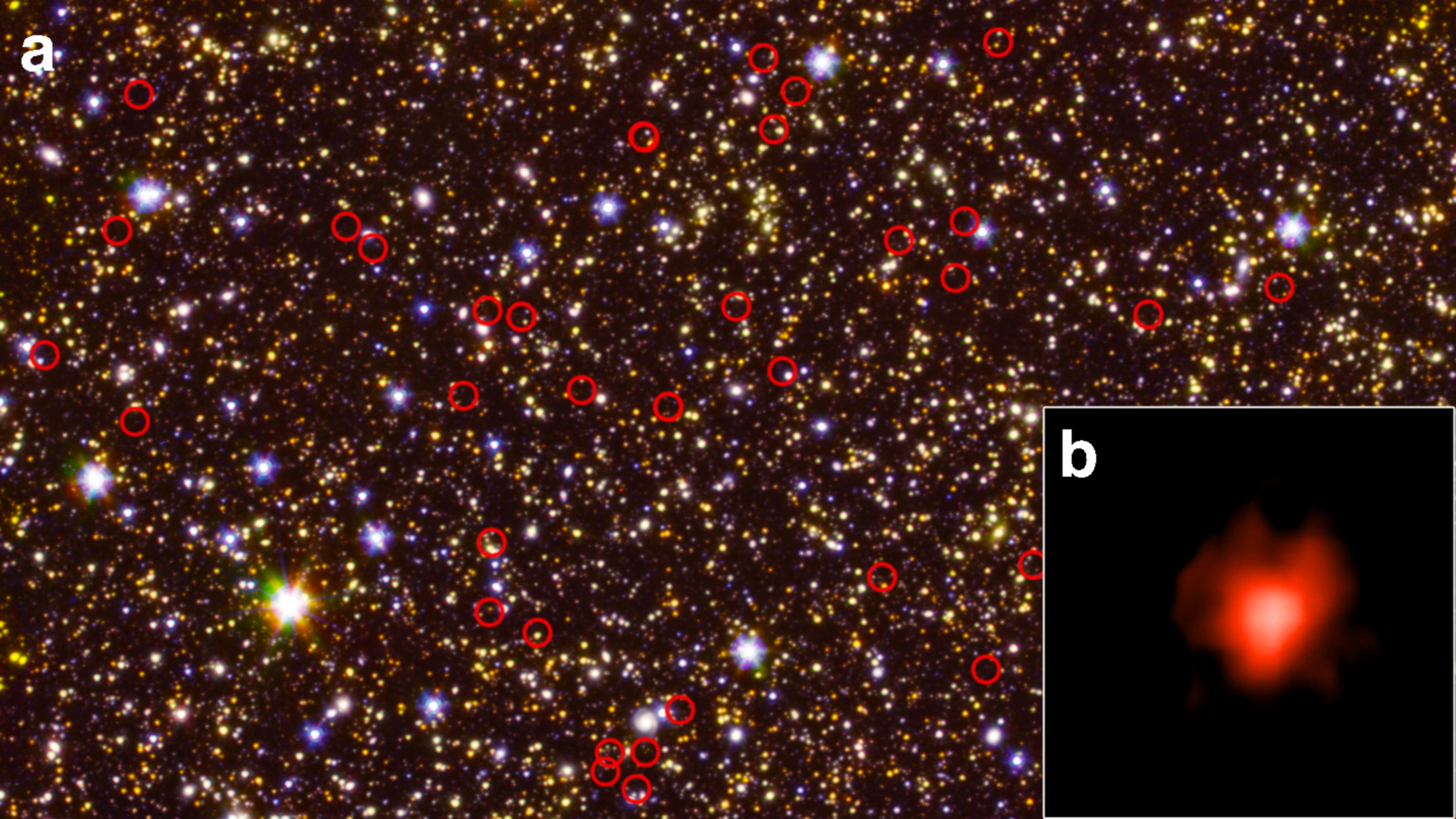}
\end{minipage}
\caption{{\bf High-redshift galaxies detected in {\Spitzer} images.}
  (a) Color composite image of the central deepest region of the
 GREATS (GOODS Re-ionization Era wide-Area Treasury). A subset of $z\sim 8$
 galaxies from \citep{debarros19} are circled in red. (b) An example of
 such a galaxy is shown in the inset. Image credit:
 NASA/JPL-Caltech/ESA/Spitzer/P. Oesch/S. De Barros/ I.Labbe. \label{fig:sudf}}
\end{figure}

Deep observations with {\Spitzer} were first undertaken in the HST deep
fields. Of the most ambitious projects with {\Spitzer} designed (in part) for the purposes of observing
high-redshift ($z>6$) galaxies was the  deep IRAC imaging of the HUDF
field (IRAC Ultradeep Field IUDF program PI
Labb\'{e}, \citealp{labbe10}). 
Many large {\Spitzer} surveys followed, 
the deepest blank-field survey to date being GREATS \citep{stefanon19}, with a
near-homogeneous observing depth of 200 hours over $\sim 200 \mbox{arcmin}^2$ (see
also Fig.~\ref{fig:sudf}). Results quickly revealed that the galaxies at
these redshifts have different rest-frame optical properties from their
lower-redshift counterparts.

From the very beginning {\Spitzer} benefited
greatly from the magnification due to gravitational lensing. While
lensing decreases the effective area surveyed, it more than
compensates by increasing the depth and resolution (Fig.~\ref{fig:surveys}).  Many surveys have recently been executed with this in mind.  Among the largest
are the Cluster Lensing And Supernova survey with Hubble (CLASH;
\citealp{postman12}), Hubble Frontier Fields (HFF;
\citealp{lotz17}), Reionization Lensing Cluster Survey (RELICS;
\citealp{coe19}). Many targets for these surveys have been selected from
Massive Cluster Survey (MACS; \citealp{ebeling07,repp18}). Each of these
surveys has its {\Spitzer} counterpart with IRAC Lensing Survey (PI
Egami), iCLASH (PI Bouwens, \citealp{bouwens14a}),
SURFSUP (PI Brada\v{c}, \citealp{surfsup}), SHFF (PI Capak), SRELICS (PI
Bradac, \citealp{strait20}). These surveys have delivered many
interesting results, especially at the lower end of the galaxy stellar mass
function. For the largest among them, catalogs have been published as
well and stellar properties investigated \citep{merlin16,castellano16,
  dicriscienzo17, santini17,shipley18,bradac19}. Perhaps the highest-redshift
galaxy so far detected by {\Spitzer} to date is a
lensed galaxy. In \citep{lam19b} the authors report a detection of MACS0647-JD $z \sim 11$ galaxy candidate strongly lensed by a cluster. It
is likely a massive and rapidly star-forming galaxy.

Already  the first samples of galaxies that were
detected  in \emph{Spitzer}/IRAC images (e.g.,
\citealp{yan06,labbe10,capak11b,labbe13}) showed
stellar masses in the range from $\sim 10^9$ to $\sim
10^{11}\,M_{\odot}$. Galaxies with stellar masses comparable to the
Milky Way masses seemed surprisingly large for a universe younger than 1
Gyr old (e.g., \citealp{yan05}). However, it quickly became clear that
at least for some of these galaxies their masses have likely been
overestimated, due to 
their IRAC fluxes being boosted by strong nebular emission lines like {\oiiif}+{\hb}  (\citealp{Finkelstein:2013fx, Smit:2014cg,
  DeBarros:2014fa,robertsborsani16, jiang16, debarros19}). 

\begin{figure}
\includegraphics[width=\textwidth]{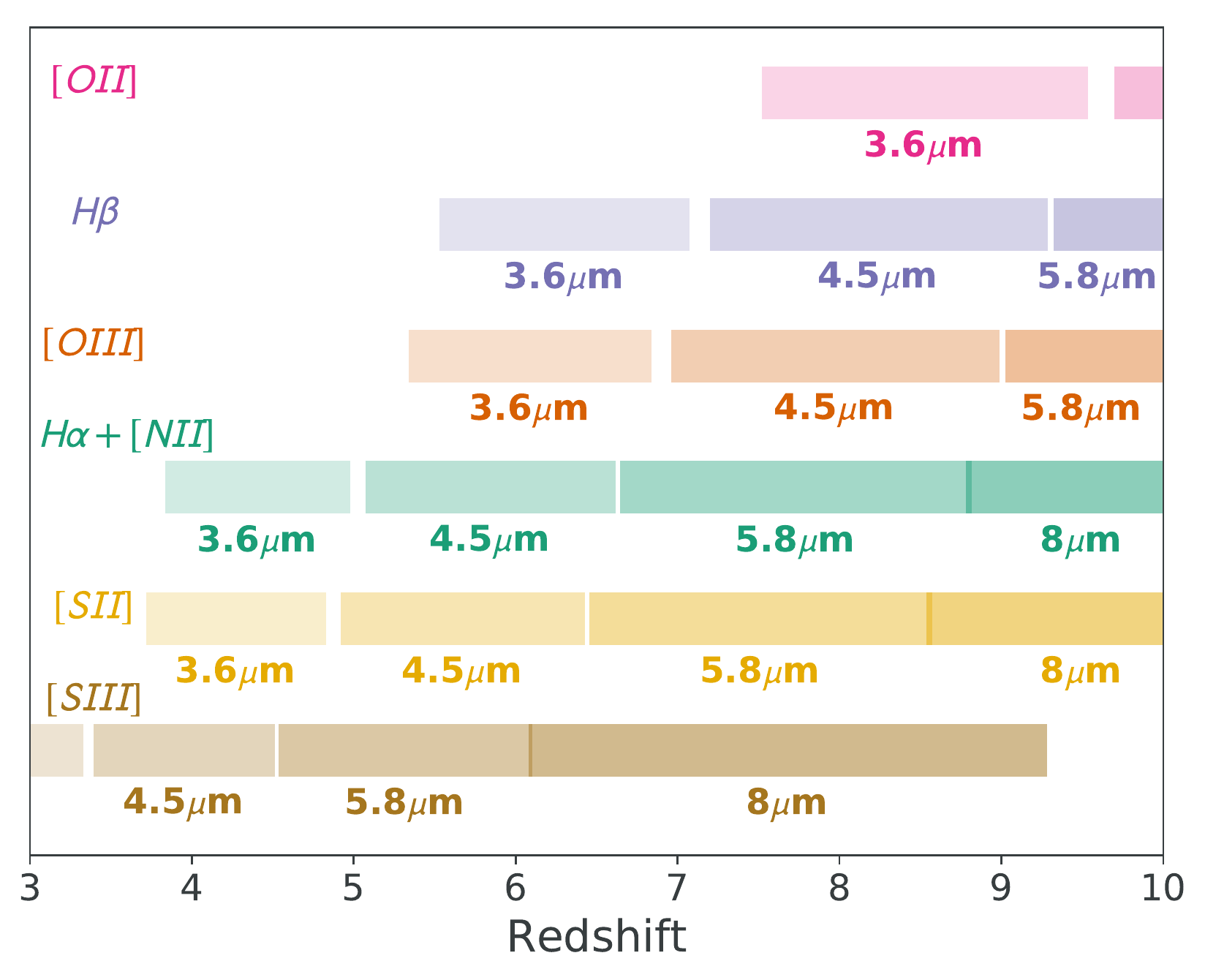} 
  \caption{{\bf Spitzer data can be used to measure (combined) strengths of
  several prominent optical lines.} Lines visible in different
{\Spitzer} channels are plotted as a function of redshift.  \label{fig:lines}}

\end{figure}

{\Spitzer} broad-band photometry can provide
indirect measurements of the nebular lines. For example, {\Spitzer}/IRAC
colors can be used to measure strengths of {\oiii}+{\hb} lines for
$6.6 \lesssim z \lesssim 6.9$ galaxies, for which these lines are
expected to fall in the ${\chone}$ band while the ${\chtwo}$ band is
relatively free of line contamination  (Fig.~\ref{fig:lines};
\citealp{shim11,smit15, robertsborsani16,huang16}). Similarly,
{\oiii}+{\hb} lines land in ${\chtwo}$ for galaxies at $7 \lesssim z
\lesssim 9$ [\citealp{laporte14,debarros19, strait20,stefanon19,
  bridge19,strait20}], while the ${\chone}$ band remains relatively free of
strong emission lines.

The results from these works show that we are now faced with a puzzle. Taken
the {\Spitzer} colors at face value, the line strengths required to fit the data for
many of these galaxies is extremely high, with rest-frame equivalent widths $EW>2000\mbox{\AA}$. But there
is also a degeneracy in these measurements, as the color can also be
boosted by an older stellar population in the
form of a Balmer break and/or dust \citep{katz19}. Thus, rest-frame
optical emission lines are likely not to be the sole cause of excess flux in the rest-frame optical bands. Unfortunately, spectroscopic observations
of rest-frame optical emission lines of the highest redshift galaxies are
out of reach for the current instruments. The sensitivity of the InfraRed Spectrograph
(IRS) on {\Spitzer} \citep{houck04} was
mostly limited to observations of $z<3$ IR luminous galaxies
\citep{teplitz07,pope08}, and we will have to wait for  JWST to allow
us to perform efficient spectroscopic follow-up in the
rest-frame optical.

However, with {\Spitzer} we are already able to study some exceptional cases where we
can mitigate or break these degeneracies. In particular, when
spectroscopic redshifts are known, we can break the
degeneracy by considering particular redshift ranges where lines are
shifted out of the filters and/or by adding external data. For
example, one can, in principle, use {\lya} fluxes to estimate the
nebular emission line contribution \citep{jiang16} or can constrain
the amount of dust using measurements using ALMA \citep{hashimoto18,laporte17}.

\begin{figure}
\begin{minipage}{\textwidth}
  \includegraphics[width=1.0\textwidth]{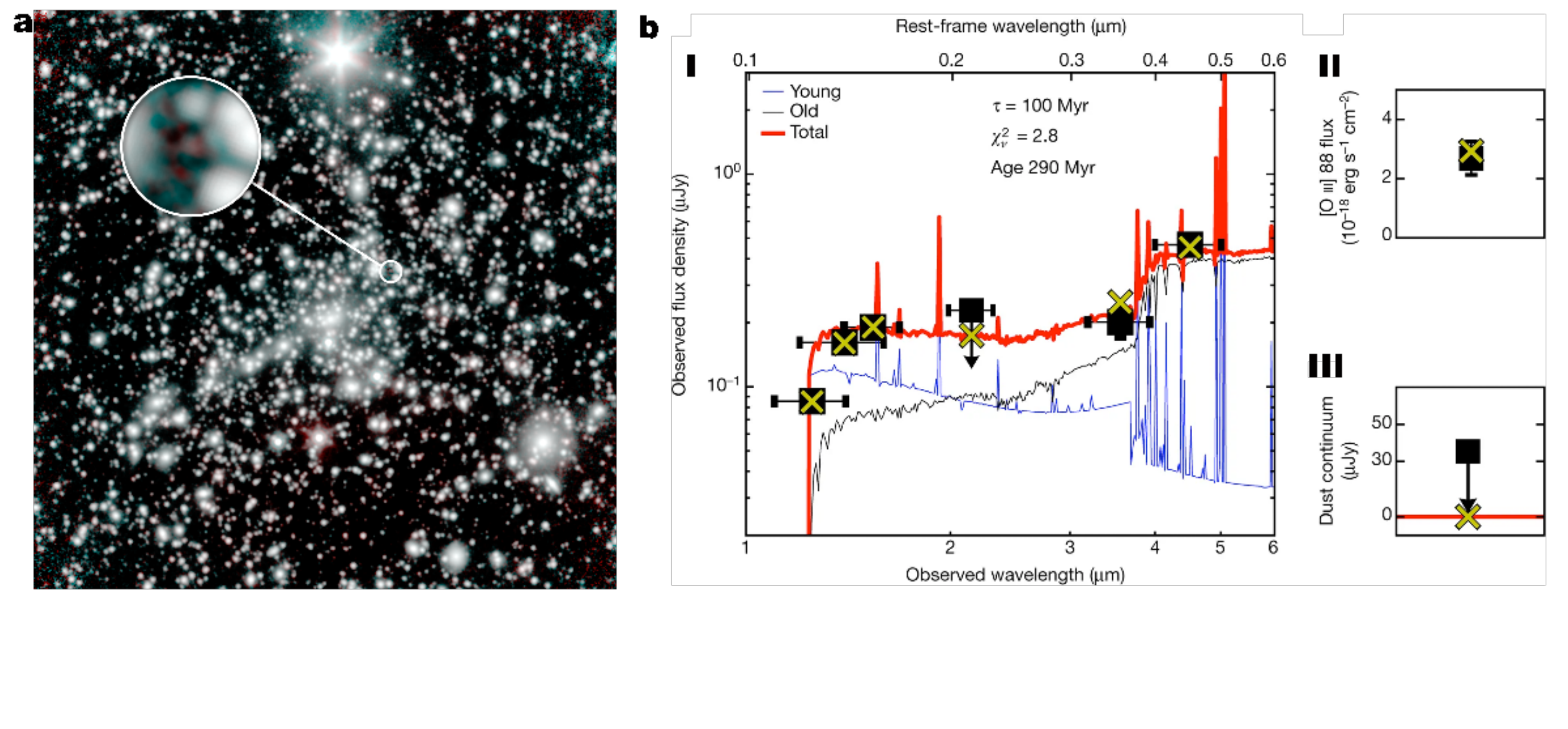}
\end{minipage}
\caption{{\bf MACS1149-JD is the best example with an evidence of old
    stellar population at $z\sim 9$.} {\bf a} {\bf Spitzer image of
    MACS1149-JD from {\surfsup} survey.} MACS1149-JD  was first discovered
  by \citep{zheng12} and first detected in both ${\chone}$
    and ${\chtwo}$ in {\surfsup} survey (
    \protect{\citealp{surfsup}}, image from {\Spitzer} press
    release feature14-13). {\bf
    b} {\bf SED modeling of the object.} SED modeling provided an evidence of old
    stellar population once the redshift was determined to be at
    $z=9.1096 \pm 0.0006$. {\bf bI} In particular, {\Spitzer} fluxes can not be
    fit without an inclusion of old ($290^{+190}_{-120}\mbox{Myr}$ old) stellar
    population.  Black squares show (from left to right) F125W, F140W and F160W data from HST, a $2\sigma$ upper limit for the Ksband from VLT/HAWK-I, and ${\chone}$
    and ${\chtwo}$  fluxes from {\Spitzer}/IRAC. The red solid line
    indicates the SED model and the corresponding magnitudes are shown
    by yellow crosses. Blue and black lines represent the
    contributions from the young ($3^{+2}_{-1}\mbox{Myr}$) and 
    old populations, respectively. {\bf bII}, The black square is the observed
    {\otf} emission line flux and its 1$\sigma$ uncertainty, while
    the yellow cross indicates the model prediction. {\bf bIII}, The black square shows the 2$\sigma$ upper limit for the dust continuum flux density, and the yellow cross indicates the model prediction. \citep{hashimoto18}. \label{fig:1149}}

\end{figure}

The best case where the degeneracy has been lifted is a
$z=9.1$ galaxy MACS1149-JD behind the cluster {\macsb}.  MACS1149-JD was originally discovered in HST and shallow
${\chtwo}$ {\Spitzer} data in \citep{zheng12}.  It was later detected
in both, $\chone$ and $\chtwo$, 
{\Spitzer} bands using deeper data
(Fig.~\ref{fig:1149}; \citealp{surfsup,huang16,zheng17}) and its
redshift was spectroscopically measured by \citep{hashimoto18}. For this galaxy the nebular emission lines are redshifted out of these
{\Spitzer} bands, yet it has a strong color excess. In
addition, the cold dust content of the galaxy was constrained to be
modest from observations taken with ALMA, making at least cold dust an unlikely
cause of the red {\Spitzer} color \citep{hashimoto18}. It is therefore
highly likely the old  ($\sim 300\mbox{Myr}$) stellar population is causing the red rest-frame optical color (\citealp{hashimoto18,hoag17,huang16}). This is surprising, given the galaxies would need to start forming significant amounts of stars shortly after the Big Bang. At $z=9.1$, when the
Universe is only $\lesssim 550 \mbox{Myr}$ old, the presence of a
strong Balmer break can thus provide the timing of the first star
formation. In the case of MACS1149-JD the dominant stellar component
formed about 250 million years after the Big Bang, corresponding to a
redshift of about 15.

There are several other objects
with {\Spitzer} detections at $z\gtrsim 9$ where this experiment can
be repeated. The highest spectroscopically confirmed galaxy detected by
{\Spitzer} (GN-z11, \citealp{oesch16}) does not show a red color.
There are others that do, e.g. GN-z10-3 and GN-z9-1 
[\citealp{oesch14}] and Abell1763-1434 [\citealp{strait20}]. They all have
${\chtwo}$ excesses that are consistent with  an evolved stellar
population, but unfortunately they currently  lack 
spectroscopic  confirmations. If their redshifts are confirmed, it is
likely that a re-analysis of these galaxies would indicate that their
star formation occurred within $300\mbox{Myr}$ of the Big
Bang.

While the sample is still small, observations of galaxies at
$z> 6$ with {\Spitzer} nonetheless show that galaxies have
unexpected properties. A large
fraction of high-redshift galaxies have either unusually strong
nebular emission lines, pronounced Balmer breaks indicating old
stellar populations, or the large amounts of dust. All three possibilities
are difficult to reconcile given the age of the Universe. E.g.,
old stellar population in MACS1149-JD puts formation redshift of the
majority of stellar population ($\sim 10^9 M_{\odot}$) at $z\sim 15$ which is
high given that simulations predict first stars only  started forming at $z\sim 30$
[\citealp{dayal18}]. Other objects are routinely showing large rest-frame
equivalent widths ($EW\sim
1000\mbox{\AA}$)  of nebular
emission lines, these have not been observed at lower redshifts except for in
some extreme cases (e.g., \citealp{du20,li18}). IRAC color excess
could also be explained by the presence of dust, but it is difficult to
produce a significant amount of dust needed to explain the
observations \citep{katz19}. While more precise answers await the James
Webb Space Telescope ({\JWST}), it is clear even at present, the star formation models used in simulations at the highest redshift are being constrained as a result of observations by {\Spitzer}.

\section{Cosmic Morning \label{sec:morning}} 
Studies of intermediate redshift ($4<z<6$) have also thrived
thanks to {\Spitzer} observations. One of the main diagnostics in
galaxy evolution models is the evolution of the Star Formation Rate
Density (SFRD, \citealp{madau14}). However, robust determination of dust attenuation is essential to transform FUV luminosity
densities into total SFRDs. Prior to {\Spitzer} observations
star formation history was determined
out to $z\sim 4$ using mostly HST data \citep{madau96,madau98,lilly96}. It was only after {\Spitzer} data
(along with Herschel) was obtained,
that the history of cosmic star formation could be robustly measured, and
the finding that the majority of star formation density comes from dust-obscured sources was established (\citealp{madau14} and references therein). MIPS also played an important role
in these discoveries at lower redshifts (e.g., \citep{lefloch05, magnelli11}).

The true power of {\Spitzer} comes with robust measurements of stellar
masses at these redshifts. The stellar-mass function measurements were first undertaken for SCANDELS, SCOSMOS, SPLASH, and UDF fields (e.g.,
\citealp{davidzon17, song16,grazian15, caputi15, salmon15,
  steinhardt14}). 
But the largest of surveys to explore this are The
Spitzer Matching Survey of the UltraVISTA Ultra-deep Stripes (SMUVS,
PI Caputi, \citealp{ashby18, caputi17}) and Euclid/WFIRST {\Spitzer}
Legacy Survey (Moneti et al. in prep.). With SMUVS, a large
fraction of galaxies  ($\gtrsim 50\%$) which were previously detected in the optical,
were, for the first time, also detected with {\Spitzer}. This
allowed for a precise measurement of the stellar mass
function. The latest results indicate that massive and
intermediate-mass galaxies have different evolutionary paths in the
early universe \citep{deshmukh18}.

One of the key properties to describe galaxy  growth is the ratio
between star formation rate (SFR) and stellar mass ($M_{*}$), also
known as specific star formation rate
($\mbox{sSFR}=\mbox{SFR}/{M_{*}}$; \citealp{guzman97,madau14}). This ratio
depends heavily on good estimates of both stellar masses and SFR.  In addition, sSFR is
particularly well-determined in the lensing fields, as its value is
independent of magnification, and yet
gravitational lensing allows us to study sub-$L^*$ galaxies at high
redshift \citep{santini17}. Its evolution with redshift is one of the
key questions in galaxy formation studies and we are still unclear as to exactly how sSFR
evolves. While pioneering studies predicted a constant sSFR
at high redshifts \citep{stark09,gonzalez10}, the later results, which included nebular emission lines in SED fitting \citep{schaerer09} and better {\Spitzer}
data find an increase at
$4<z<6$ and beyond (e.g.,
\citealp{davidzon18,santini17,salmon15,grazian15,
  tasca15, debarros14,
  duncan14, gonzalez14, stark13}).

With {\Spitzer} also SFR using {\ha} was investigated in detail
for the first time  at $4<z<5$, which is regarded as one of the most
reliable
among the easily accessible nebular SFR tracers
(e.g.,\citealp{moustakas06}).  Just like with {\oiii}, at intermediate
redshifts there are ranges where 
emission line strength of {\ha} can be measured with {\Spitzer} (\citealp{shim11},  Fig.~\ref{fig:lines}). One of the caveats is that
{\ha} is blended with
{\nii}. Still, this has been used to
indirectly measure the {\ha} strength at $3.9\lesssim z \lesssim 4.9$ and
deduce the star formation rates \citep{stark13,salmon15, smit14,
  bouwens16b, marmolqueralto16,faisst16,caputi17}. More
recently, in \citep{faisst19} authors measured burstiness of star formation by
comparing ultra-violet, H$\alpha$ luminosity, and
H$\alpha$ equivalent-width  of $z\sim 4-5$ main sequence galaxies,
indicating that for at least half of their sample the star formation
history is not smooth.  Access to {\ha} also allows modelling of 
Lyman-continuum photon production efficiency $\xi_{\rm
  ion}$. In \citep{lam19} the authors used stacking of 300 $3 < z < 6$ galaxies to estimate {\ha} equivalent
widths and found that $\xi_{\rm
  ion}$ is not strongly dependent on luminosity and is thus similar to
 the values derived for brighter galaxies. This has important
 implications, in particular for establishing that
 faint galaxies are able to produce the Lyman-continuum photons needed for cosmic reionization.

 While {\Spitzer} has improved our knowledge of stellar masses and
 SFRs considerably, several issues remain. In particular, the
 evolution of sSFR is still not a completely solved problem. The
 situation will likely improve soon with the data taken from
 {\Spitzer} in the last year, as well as in the future with missions
 like {\JWST}, Euclid and the Wide-field Infrared Survey Telescope
 (WFIRST).

\section{Cosmic Dusk} 
The present epoch is one of a rapid decline in the global star
formation rate, but clusters experience an evolution
in star formation activity over this time that is even stronger than in
the field. Identifying the processes that trigger and
terminate star formation in cluster galaxies
(e.g. ram-pressure stripping, starvation, merging, harassment; \citealp{dressler99,
poggianti99}), and contrasting them to those operating in the field is
key to understanding the causes of the general decline. Furthermore,
studying clusters at earlier times provides new constraints on both
the evolution of the cluster abundance and the evolution of early-type
galaxies over a substantial look-back time \citep{rosati02}. {\Spitzer}
plays a large role in identifying both high-redshift ($z>1$) clusters
and protoclusters. Red-sequence overdensities and color searches can efficiently  be done with
{\Spitzer} and have been utilized in both high-redshift cluster and
protocluster searches. In particular, the inclusion of {\Spitzer} data
dramatically increases the accuracy/precision of photometric redshifts
at $z>1$, which is crucial for finding structures and to robustly
estimate the environment (e.g., \citealp{lemaux18}).

\begin{figure}
\includegraphics[width=1\textwidth]{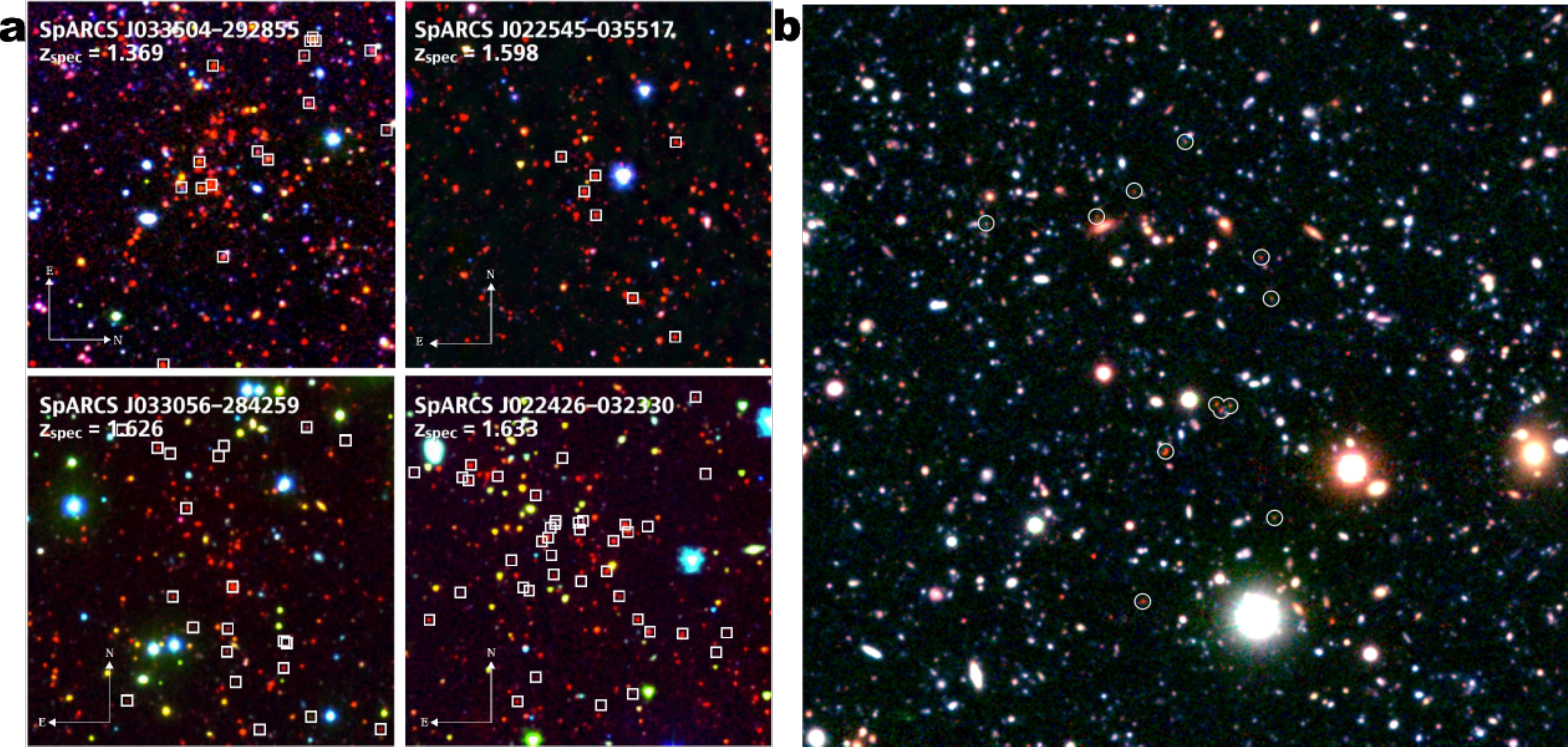} 
  \caption{{\bf Spitzer data has been very effective in searching for and
    characterizing $z>1$ galaxy clusters and protoclusters.}
{\bf a} {\bf Examples of high-redshift clusters  from SpARCS survey}. Tri-color gz[3.6] images of the
    central regions of the four southern $z> 1.35$ SpARCS
    clusters. Spectroscopic cluster members are marked with white
    squares \citep{nantais16}. {\bf b} {\bf Image of a proto-cluster
    of galaxies at $z=5.3$.} The cluster was discovered by a suite of
    multi-wavelength observations made with {\Spitzer}, Chandra, HST,
    Subaru and Keck telescopes. Members of the developing cluster are
    circled in white (Spitzer press release ssc2011, \citealp{capak11}).}
\label{fig:sparcs}
\end{figure}

Early on, {\Spitzer} helped in discoveries of the protoclusters. In
\citep{capak11} the authors present the discovery of a $z=5.7$
protocluster that dates back to $1\mbox{Gyr}$ after the Big Bang (Fig.~\ref{fig:sparcs}). The field
contains a luminous quasar as well as a large galaxy rich in molecular
gas. The Clusters Around Radio-Loud AGN (CARLA) survey used this fact,
as CARLA targeted radio-loud quasars with {\Spitzer} at $z>1.3$,
discovering several proto-clusters surrounding them \citep{galametz12,
  noirot18}.  {\Spitzer} data is used to improve redshift information,
as well as to estimate stellar masses, SFR and ages as these are the only
bands available that probe the optical/NIR rest-frame for
high-redshift protoclusters.

{\Spitzer} played an even more prominent role in finding the highest redshift galaxy clusters. Shallow yet
wide surveys, in particular the IRAC Shallow Survey, enabled detections
of many high-redshift clusters
(\citealp{stanford05,stanford06,eisenhardt08}).  More discoveries were
made in the {\Spitzer} Wide-Area Infrared
Extragalactic (SWIRE) Survey using a simple color excess in {\Spitzer}
bands \citep{papovich10}. The large area is needed as clusters of galaxies are
extremely rare, and one requires such a survey to find the most massive
examples (using numbers from \citep{gonzalez19} $\sim 6~\mbox{deg}^2$ are
needed to detect one high-redshift $z\gtrsim 1$, massive
$M>10^{14}M_{\odot}$ cluster; Fig.~\ref{fig:surveys}).

The {\Spitzer} Adaptation of the
Red-sequence Cluster Survey (SPARCS; \citealp{wilson09,muzzin09}; Fig.~\ref{fig:sparcs}) was
the first to  perform a comprehensive study of $z>1$
galaxy clusters with all IRAC and MIPS bands.  At $z=1.5$, the 1.6
micron bump \citep{sawicki02}, due to infrared emission of stars, is
shifted into {\Spitzer} ${\chtwo}$ band, making it a useful tool to
detect cluster members. Finally,
surveys like IRAC Distant Cluster Survey \citep{stanford12}, MaDCoWS
(using WISE, but utilizing {\Spitzer} for confirmation;
\citealp{gonzalez19}) and {\Spitzer} South Pole Telescope Deep Field
\citep{rettura14,ashby13} have also detected many rich clusters at
$z>1$.   Even more importantly, both IRAC and MIPS
enabled the study the stellar properties of
cluster members \citep{brodwin13,
  nantais16}. The main conclusion from these surveys is that while low-redshift
clusters are considered mostly star formation graveyards, at the
earlier times ($z>1$), galaxies in galaxy clusters were star forming and
active \citep{brodwin13}. This is thus the star-forming epoch of galaxy clusters, the
study of which was enabled by {\Spitzer}.

\section{Thank You Spitzer}
When {\Spitzer} was first planned, nobody was expecting it to do the
groundbreaking discoveries at the cosmic dawn. This was mostly due to
the mirror size, yet the 85cm diameter mirror (only slightly larger
than a wine-barrel) has
surpassed a lot of predictions. It was meant to explore ``the Old, the
Cold, and the Dusty'', and yet it also explored the starry, the
many, and the first. {\Spitzer} let us study early star formation in the first galaxies at the
epoch of reionization, it delivered new insights on star formation at
high-redshift, and allowed us to find and study the earliest galaxy clusters
and protoclusters. While the next space telescope (JWST) will revolutionize these fields in many ways,
it will not surpass {\Spitzer} in its ability to do survey science
thanks to {\Spitzer}'s combination of a large field of view,
excellent sensitivity, and long life. So long {\Spitzer}, and thanks for all the
photons.

\begin{addendum}
 \item Based on observations made with the  {\Spitzer} Space Telescope, which is operated by the
Jet Propulsion Laboratory, California Institute of Technology under a
contract with NASA. Support for this work was provided by NASA
through ADAP grant 80NSSC18K0945, NSF grant AST 1815458 and through
an award issued by JPL/Caltech. The author would like to thank Brian
Lemaux and Victoria Strait for their help with the manuscript.
\item[Financial Interest] The authors declare no competing financial interests.
\item[Correspondence] Correspondence should be addressed to M.B.
\end{addendum}

\bibliography{bibliogr_cv,bibliogr_highz}

\end{document}